# Multifractality, percolation threshold and critical point of a nuclear reactor

V. V. Ryazanov

Institute for Nuclear Research, pr. Nauki, 47 Kiev, Ukraine, e-mail: vryazan19@gmail.com

A multifractal model is used to analyze neutron evolution within a reactor. For chain reactions, various characteristics of multifractal neutron behavior have been determined. These include the dimension of the multifractal carrier, information and correlation dimensions, the entropy of the fractal set, maximum and minimum dimension values, and the multifractal spectrum function. The geometric features of a multifractal allow for the description of a stochastic system consisting of hierarchically subordinate statistical ensembles, which are characterized by Cayley trees. A stationary distribution over hierarchical levels is established, which follows the Tsallis power law. The text also points out some potential applications of fractal patterns in nuclear reactor theory. The chance of percolation, which is when we see a state in the Bethe lattice where there's at least one continuous path through neighboring conducting nodes all the way across, is similar to the likelihood of a self-sustaining fission chain reaction happening. When this probability hits a critical point, we get a (conditionally) infinite cluster of neutrons forming. The percolation probability, influenced by how long the reactor has been running and its size, is linked to the reactor's criticality. We take a look at how the neutron multiplication factor behaves over time. We especially focus on the early stages of a self-sustaining nuclear fission chain reaction. We also highlight the ways to identify the boundaries of the critical region.

## 1. Introduction

Fractals with variable dimensions, also known as multifractals, are actually more common in nature than those exact self-similar fractals with constant dimensions, which we rarely see in everyday life. A multifractal is a unique, non-uniform fractal object. When we look at fractal sets that represent real physical formations and processes, we find that they usually don't just have one dimension; instead, they have a whole spectrum of dimensions! Besides the geometric traits we find in monofractals with a single dimension, these sets also come with some interesting statistical properties. Take neutron fission chains in a reactor's chain reaction, for instance—they fit right into this fascinating category!

Following the ideas from [1, 2], let's talk about the fractal dimension! The space where our fractal object hangs out is known as the embedding space. It's just your typical Euclidean space with dimension $d$. Now, imagine wrapping the whole fractal object in $d$-dimensional "balls" that have a radius of $l$. If this required at least $N(l)$ balls, and for sufficiently small $l$, the value of $N(l)$ changes with $l$ according to a power law

$$N(l) \sim l^{-D},$$

then $D$ is called the Hausdorff or fractal dimension of this object [1, 2].

A multifractal is fascinating because it has a unique spectrum! Its dimension changes based on the non-additivity parameter $q$, which tells us how diverse the multifractal is. This parameter $q$ pops up in non-additive thermodynamics and in various statistical distributions, like the Renyi and Tsallis distributions [3]. It shows us how much things deviate from being additive. When $q$ is equal to 1, thermodynamics turns additive, and the Tsallis distribution aligns with the Gibbsian one. When it comes to natural fractals, there's a certain scale where the cool feature of self-similarity just fades away. For neutrons hanging out in a reactor, this happens around the neutron free path $\lambda$. Self-similarity also takes a backseat at really large scales. In the case of reactors, we're talking about correlation lengths, which are the typical spatial scale of the cluster, linked to the critical size of the reactor when it's close to that critical point. About how to describe fractals on finite scales, you can check out the discussion in [4].

The exploration of multifractality and the potential for its numerical characterization is an ascent technique for examining empirical data, relevant to abroad spectrum of issues. In [5-7], multifractal examination of intricate signals is detailed, applicable, for instance, to fluctuating reactor power. Employing multifractal fluctuation assessment, time series in finance, health care, and weather forecasting were investigated. It may be utilized with equal effectiveness in diverse challenges with in nuclear reactor theory.

The aspects brought up in this study possess many other varied facets. Hence, employing geometric properties of a multifractal permits the depiction of a random system comprising statistically subordinate ensembles arranged hierarchically and defined by Cayley trees [6-8], such as neutron propagation in a reactor. In [6, 7], it is demonstrated that the development of hierarchical frameworks simplifies to anomalous diffusion with in ultrametric space, leading to a steady distribution across hierarchical tiers that corresponds to the Tsallis power law typical of non-additive systems [3, 9]. The Tsallis distribution is a particular instance of superstatistics [10], connected to the distribution encompassing the lifespan [11], which was utilized to examine procedures in a nuclear reactor [12, 13].

The progression of neutrons in the reactor follows fractal paths similar to Cayley trees. This is verified by the successful portrayal of neutron events in a reactor through stochastic branching mechanisms [14, 15]. Much of the information gained for fractal frameworks can be utilized in the study of nuclear reactors. Specifically, the non-uniform Sierpinski triangle may function as an efficient representation for characterizing chain reactions within a reactor.

Anomalous diffusion, along with such properties as the time of initial level achievement, etc., on the Sierpinski triangle and branching configurations are examined, for instance, in [16]. Alikeness can be drawn with statistical mechanics, where the Ising model acts as a potent framework. Reactor attributes such as critical dimension, output, disturbance spread velocity, etc., are linked to fractal outlines. Consequently, incorporating fractal designs when depicting neutron conduct in a reactor appears valuable and essential. Simultaneously, numerous established occurrences are investigated from an alternate view point, enabling the uncovering of some novel aspects of already familiar matters. Numerous findings from the theory of fractals may find utility in reactor theory.

The multifractal approach has proven effective when applied across various physical, chemical, and biological challenges. It is reasonable to assume that exploring these characteristics of neutron behavior within a nuclear reactor may provide deeper insights into issues critical to reactor safety. For instance, it could help refine the understanding of how the critical point's position depends on system size, determine the reactor's critical dimensions, and analyze the rate at which disturbances propagate within the reactor. In a related study [17], the author compares the phenomenon of percolation, which he explores, to a chain reaction, noting similarities between the spread of rumors in the percolation model and the explosion of an atomic bomb. Furthermore, the connection between the fractal geometry of Cayley trees and neutron multiplication processes in reactors can be supported by their shared foundation in stochastic branching mechanisms, such as Galton-Watson processes [14, 15].

Neutron trajectories within a reactor exhibit real-world fractal structures resembling Cayley trees. Consequently, the study of neutron processes in reactors should incorporate methods grounded in the extensive science of fractals, alongside traditional approaches. Many other fractal patterns display self-similarity, such as velocity fields in turbulent fluid flows, lightning discharges, cloud formations, the human circulatory system, tree contours, and more. Multifractal analysis has proven effective in various areas, including the structural distribution of inhomogeneous star clusters in astrophysics, examining aggregation properties of blood cell elements in biology, characterizing key stages of dislocation evolution and material fatigue fractures in metal physics. Moreover, multifractal concepts find applications in diverse fields such as developed hydrodynamic turbulence, analyzing incommensurate structures and quasicrystals in solid-state physics, studying spin glasses and disordered systems, as well as in quantum mechanics and elementary particle physics [2].

In reference [18], the strict correlations within the theory of percolation on Bethe lattices are utilized to describe the behavior of the neutron multiplication factor. The reactor's critical point aligns with the percolation threshold. The analysis includes the percolation probability, representing the likelihood of a self-sustaining chain reaction, along with its derivatives. A notable feature that underscores the complexity and nonequilibrium dynamics of nuclear chain reactions within a reactor is their hierarchical organization.

This study employs the statistical framework of hierarchical systems to conduct a more in-depth examination of intricate fission chains.

Hierarchical subordination concepts have been widely applied to explain various complex systems, including physical, biological, economic, environmental, and social systems, among others. One of the most promising uses of hierarchical structure theory is found in the study of complex networks [19]. Real-world networks tend to exhibit significant clustering and a self-similar structure, often characterized by a power-law probability distribution that governs the number of connections between neighboring nodes [20, 21]. Additionally, many networks display a block structure, enabling the identification of node groups that are densely interconnected within themselves but maintain weak or no connections with external nodes outside the group. This phenomenon occurs because, in systems far from equilibrium where ergodicity is lost, phase space fragments into clusters aligned with structural levels hierarchically organized. A similar behavior is observed in fission chains within nuclear reactors. Such hierarchically subordinate systems are represented within an ultrametric space [20, 22-23, 8], which can be geometrically visualized through a Cayley tree diagram (Fig. 1).

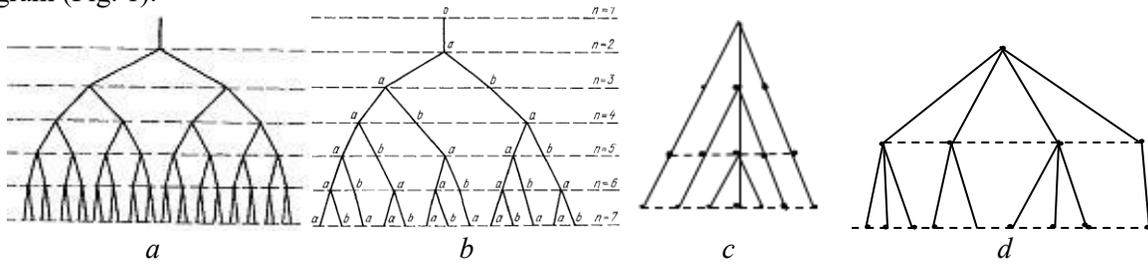

Fig.1. *a*). The simplest regular Cayley tree with branching s=2; b). Irregular Fibonacci tree with variable branching; c). Degenerate tree with s=3; d). Irregular tree for *n*=2, *a*=2.

This paper applies results from the theory of hierarchically subordinate systems far from equilibrium to describe nuclear chain reactions within nuclear reactors. A connection is identified between the percolation characteristics of neutron-nuclear processes, discussed in [18], and the hierarchical intensity at level *n*. For stochastic systems, this intensity corresponds to the probability density and relates to the degree of hierarchical linkage, *w*, associated with the tree nodes at a given level. Explicit expressions for these quantities are derived for various hierarchical tree structures, establishing their correspondence with different reactor operating modes and neutron trajectories under these conditions. The study also examines processes of anomalous diffusion in ultrametric spaces, presenting stationary solutions in the form of the Tsallis distribution [24]. It is demonstrated that, in specific cases, these distributions correlate with Renyi distributions, which are linked to percolation probabilities and the likelihood of fission chain reactions.

The article is organized as follows. Section 2 examines the multifractal properties of neutron processes in a reactor. A spectrum of generalized dimensions and a multifractal spectrum function are obtained for fission chains in nuclear reactors, taking into account delayed neutrons. Section 3 applies the concept of hierarchical subordination to chain reactions in a nuclear reactor. Section 4 examines the relationship between the percolation threshold and the critical point of a nuclear reactor. This discussion continues in Section 5, which explores the relationship between the theory of percolation on Bethe lattices and the neutron multiplication factor. A discussion of the results and final conclusions are presented in Section 6.

## 2. Neutron transfer processes in the reactor and multifractality of fission

To determine the distribution of the relative density of the number of neutrons, as in [1, 2], we divide the entire reactor region of size *L* into cubic cells with a side of size $\varepsilon_0 = L/(L/\varepsilon_0) = l_0 \ll L$ ($L/\varepsilon_0$ is the number of cells) and volume $\varepsilon_0^d$, where $l_0$ is of the order of the neutron mean free path $\lambda$, *d* is the Euclidean dimension of space. The neutron population, assuming that the initial neutron is one, consists of *N* neutrons distributed throughout the reactor volume. In the limit of an infinite multiplying medium $N \to \infty$. Following the procedure applied in [1, 2] to the inhomogeneous Cantor set, the inhomogeneous Sierpinski triangle and other fractal objects, we assume that at the *n*-th step, for the *n*-th generation of neutrons, the number of cells changes as $(L/\varepsilon_0)^n$, and the cell size becomes equal to $\varepsilon_n = L/(L/\varepsilon_0)^n$.

The exponential growth in the number of cells arises because, in Cayley trees associated with division processes, both the number of nodes and potential events expand exponentially. Let the cells be represented by the index $i = 0, 1, 2, …, N - 1$. The distribution of the neutron population across the reactor volume is described by a series of values $n_i$, which represent the number of neutrons present in the *i*-th cell:

$$p_i = n_i/N = p_i(\varepsilon) = \lim_{N \to \infty} n_i(\varepsilon)/N. \tag{1}$$

This denotes the likelihood of finding a randomly chosen neutron within the *i*-th cell. In practical scenarios, the number of neutrons, $N$, is always finite, though typically quite large. The size of each cell corresponds to $\varepsilon_0 = L/(L/\varepsilon_0) = l_0$ only at the zero step, before the start of the division process corresponding to the procedure for constructing a fractal set - the Cayley tree. The cell size in the *n*-th generation of neutrons, as indicated above, is equal to $\varepsilon_n = L/(L/\varepsilon_0)^n$ and tends to zero as $n \to \infty$ (although in real reactors the values of *n* are finite, but can be very large).

To begin, let's examine a straightforward model that does not differentiate between prompt and delayed neutrons. This approach utilizes an approximation derived from branching process theory [14, 15], where a single nuclear fission event results in the production of either zero neutrons with probability $\pi_0$ or 2 neutrons with probability $\pi_2$, $\pi_0 + \pi_2 = 1$. With probability one neutron will collide with a fuel nucleus and split it ($\lambda_f = v\Sigma_f$, $v$ - neutron speed, $\Sigma_f$ - macroscopic fission cross section, $\lambda_f$ - fission intensity, the absorption intensity $\lambda_c$ is determined similarly). The value $p$ is calculated with known values of the cross sections $\sigma_f$, $\sigma_c$. The relations $k_{eff} = p\bar{v}$ are valid, where $k_{eff}$ is the effective neutron multiplication factor, $\bar{v}$ is the average number of neutrons produced during one fission of a fuel nucleus. Note that the values $p_i(\varepsilon)$ do not coincide. Thus, after the first division, the probability that in a cell with size $\varepsilon_2 = L/(L/\varepsilon_0)^2$ there are 0, 1 and 2 neutrons are equal to: $p_0(\varepsilon) = p\pi_0$, $p_1(\varepsilon) = 1-p$, $p_2(\varepsilon) = p\pi_2$. The probabilities then grow more intricate. The total of the probabilities for all possible events following a single fission becomes increasingly complex

$$\Phi = p(\pi_0 + \pi_2) + 1 - p; \quad \Phi = 1; \quad (\pi_0 + \pi_2 = 1). \tag{2}$$

In relations (1) and (2), the probability normalization condition $\sum_{i=1}^{N} p_i(\varepsilon) = 1$, $\Phi = 1$ is satisfied. The generalized partition function is defined by exponent $q$, $-\infty < q < \infty$ [2]

$$Z(q,\varepsilon) = \sum_{i=1}^{N} p_i^q(\varepsilon). \tag{3}$$

The range of generalized fractal dimensions is described in [2] through the relationship

$$D_q = \frac{\tau(q)}{q-1}, \tag{4}$$

where

$$\tau(q) = \lim_{\varepsilon \to 0} \frac{\ln Z(q,\varepsilon)}{\ln \varepsilon}. \tag{5}$$

When $D_q = D = const$ we obtain the usual regular monofractal.

To analyze the division process, we apply the algorithm outlined in [2]. When the process follows a binomial distribution, the generalized partition function, constructed as per definition (3), takes the following form for our multifractal

$$Z(q,\varepsilon) = (p_0^q + p_1^q + p_2^q)^n = \Phi_q^n, \quad \Phi_q = p^q(\pi_0^q + \pi_2^q) + (1-p)^q. \tag{6}$$

Two potentially generated neutrons have the ability to trigger subsequent events, with their combined probabilities adding up $\Phi^2$. In the second generation the probabilities are $\Phi^3$, in the third – $\Phi^4$, in the *n*-th – $\Phi^n$. From formulas (3), (5) and (6) we obtain that for $\varepsilon_n \to 0$, $n \to \infty$,

$$Z(q,\varepsilon) = \sum_{i=1}^{N} p_i^q(\varepsilon) \approx \varepsilon^{\tau(q)}. \tag{7}$$

From relations (4) - (7) for $Z(q,\varepsilon) = \Phi_q^n$, $\Phi_{q=1} = 1$ and $\varepsilon_n = L/(L/\varepsilon_0)^n$ we obtain for large *n*:

$$\tau(q) = (q-1)D_q = \frac{\ln \Phi_q}{\ln(L/\varepsilon_n) - \ln(L/\varepsilon_0)} \simeq \frac{\ln \Phi_q}{\ln(L/\varepsilon_0)}. \tag{8}$$

By setting $q = 0$ in expression (8), we determine the dimension

$$D_0 = -\frac{\ln 3}{\ln(l_0 / L)}. \tag{9}$$

This refers to the dimension of the multifractal carrier [2]. At $q = 1$

$$D_1 = -\frac{\ln S}{\ln(l_0 / L)}, \quad -S = \sum_{i=1}^{N} p_i \ln p_i = p\pi_0 \ln(p\pi_0) + p\pi_2 \ln(p\pi_2) + (1-p)\ln(1-p). \tag{10}$$

The entropy, denoted as $S$, represents the entropy of the partition of the measure on the set $L$. It also corresponds to the entropy associated with the fractal set. At $q = 2$

$$D_2 = \frac{\ln[p^2(\pi_0^2 + \pi_2^2) + (1-p)^2]}{\ln(l_0 / L)}. \tag{11}$$

Having established the definition of the pair correlation integral, $I(l) = \lim_{N \to \infty} \frac{1}{N^2} \sum_{n,m}^{N} \theta(l - |r_n - r_m|)$
The summation is performed over all pairs of points within the fractal set, considering their corresponding radius vectors $r_n$ and $r_m$, $\theta(x)=1$, if $x \geq 0$, $\theta(x)=0$, if $x < 0$, we obtain that $I(l) \approx \sum_{i=1}^{N} p_i^2 \approx (\frac{l}{L})^{D_2}$.

By introducing the probability density, $\rho(r) = \frac{1}{N} \sum_i \delta(r - r_i)$, $\int_L \rho(r) d^d r = 1$, the pair correlation function, representing the probability density of two arbitrary points in the set being separated by a distance $r$, is given by $C(r) = \int_L \rho(r')\rho(r'+r)d^d r'$. The correlation function $C(r)$ exhibits power-law behavior as a function of distance $r$, i.e. $C(r) \approx 1/r^\beta$, $\beta = d - D_2$, where $d$ is the Euclidean dimension of space, in our case, is equal to 3. The Fourier component $C(k)$, which is dependent on the wave vector $k$, varies in accordance with a power law $C(k) = 1/k^{D_2}$.

The maximum dimension value is

$$D_{\max} = D_{-\infty} = \frac{\ln(p\pi_0)}{\ln(l_0 / L)}, \tag{12}$$

its value $D_q$ reaches at $q \to -\infty$. Minimum dimension value is

$$D_{\min} = D_{\infty} = \frac{\ln(1-p)}{\ln(l_0 / L)}, \tag{13}$$

is achieved at $q \to \infty$. In formulas (12) and (13), we assumed, consistent with the fact that $k_{ef}=p\bar{v}$, with $k_{ef} \simeq 1$, $\bar{v} = 2,4$, $p \simeq 1/\bar{v} \simeq 0,4$, and put $\pi_2=0,8$, $\pi_0=0,2$.

Alongside the generalized dimensions $D_q$, the multifractal spectrum function, $f(\alpha)$ (also known as the spectrum of multifractal singularities), is utilized to describe the characteristics of a multifractal set. The probabilities $p_i$ (1), which represent the relative distribution of cells $\varepsilon$ that cover the set, exhibit a power-law behavior based on the size of the cell $\varepsilon$, particularly for self-similar sets:

$$p_i(\varepsilon) \simeq \varepsilon^{\alpha_i}, \tag{14}$$

where $\alpha_i$ represents the exponent. The $\alpha$ values span across the interval $(\alpha_{min}, \alpha_{max})$,

$$p_{\min} \simeq \varepsilon^{\alpha_{\max}}, \quad p_{\max} \simeq \varepsilon^{\alpha_{\min}}; \quad \frac{d\tau}{dq}\Big|_{q \to +\infty} = D_{\infty} = \alpha_{\min}; \quad \frac{d\tau}{dq}\Big|_{q \to -\infty} = D_{-\infty} = \alpha_{\max}. \tag{15}$$

Knowing $D_q$, the dependence $\alpha(q)$ can be determined using the equation

$$\alpha(q) = \frac{d\tau}{dq}[(q-1)D_q]. \tag{16}$$

For $\Phi_q$ (2), expression (16) assumes the following form:

$$\alpha(q) = \frac{1}{\ln(l_0 / L)} \frac{1}{\Phi_q} \frac{d\Phi_q}{dq}; \quad \frac{d\Phi_q}{dq} = (\ln p) p^q (\pi_0^q + \pi_2^q) + p^q (\pi_0^q \ln \pi_0 + \pi_2^q \ln \pi_2) + (1-p)^q \ln(1-p).$$

From the expression

$$\tau(q) = q\alpha(q) - f(\alpha(q)); \quad f(\alpha(q)) = \frac{q}{\ln(l_0 / L)\Phi_q} \frac{d\Phi_q}{dq} - \frac{1}{\ln(l_0 / L)} \ln \Phi_q, \tag{17}$$

the dependence $f(\alpha)$ is established in parametric form. Specifically, using formula (16), we determine $q(\alpha)$ and then substitute it into expression (17). The variables $\{q, \tau(q)\}$ are related to the variables $\{\alpha, f(\alpha)\}$ by the Legendre transform $\alpha = \frac{d\tau}{dq}; \quad f(\alpha) = q\frac{d\tau}{dq} - \tau$. Inverse Legendre transform is:

$$q = \frac{df}{d\alpha}; \quad \tau(q) = \alpha \frac{df}{d\alpha} - f; \quad \frac{d^2 f}{d\alpha^2} = (\frac{d^2\tau}{dq^2})^{-1}.$$

At the point $\alpha_0 = \alpha(0)$, the function $f(\alpha)$, which is convex throughout, reaches its maximum. For $\Phi_q$ (2)

$$\alpha_0 = \alpha(q=0) = \frac{q}{\ln(l_0 / L)} \frac{1}{3}[2\ln p + \ln \pi_0 + \ln \pi_2 + \ln(1-p)].$$

The fractal dimension of the measure support corresponds to $f(\alpha_0) = D_0$. Near its maximum, the function $f(\alpha)$ can be approximated by a parabola. This is due to the fact that:

$$\tau'(0) = \alpha_0; \quad f''(\alpha) = 1/\tau''(0); \quad \tau''(0) = 2(D_0 - \alpha_0) - D''_{q=0}, \quad f(\alpha) \simeq D_0 - \frac{(\alpha - \alpha_0)^2}{2[2(\alpha_0 - D_0) + D''_{q=0}]}.$$

For $\Phi_q$ from expression (2)

$$D_q = \frac{1}{\ln(l_0 / L)\Phi_q} \ln \Phi_q; \quad \frac{dD_q}{dq} = \frac{1}{\ln(l_0 / L)} (-\frac{\ln \Phi_q}{(q-1)^2} + \frac{1}{(q-1)\Phi_q} \frac{d\Phi_q}{dq});$$

$$\frac{d^2 D_q}{dq^2} = \frac{1}{\ln(l_0 / L)} \{\frac{2 \ln \Phi_q}{(q-1)^3} - \frac{2}{(q-1)^2 \Phi_q} \frac{d\Phi_q}{dq} + \frac{1}{(q-1)}[\frac{1}{\Phi_q} \frac{d^2 \Phi_q}{dq^2} - (\frac{1}{\Phi_q} \frac{d\Phi_q}{dq})^2]\};$$

$$\frac{d^2 \Phi_q}{dq^2} = (\ln p)^2 p^q (\pi_0^q + \pi_2^q) + 2(\ln p) p^q (\pi_0^q \ln \pi_0 + \pi_2^q \ln \pi_2) + p^q [\pi_0^q (\ln \pi_0)^2 + \pi_2^q (\ln \pi_2)^2] + (1-p)^q (\ln(1-p))^2;$$

$$\frac{d^2 D_q}{dq^2}\bigg|_{q=0} = D''_{q=0} = \frac{1}{\ln(l_0 / L)} \{-2\ln 3 - \frac{2}{3}[2\ln p + \ln \pi_0 + \ln \pi_2 + \ln(1-p)] + [\frac{1}{3}(2\ln p + \ln \pi_0 + \ln \pi_2 + \ln(1-p))]^2 -$$

$$-\frac{1}{3}[2(\ln p)^2 + 2\ln p(\ln \pi_0 + \ln \pi_2) + (\ln \pi_0)^2 + (\ln \pi_2)^2 + (\ln(1-p))^2]\}.$$

When $q=1$, $\alpha(1)=D_1=f(\alpha(1))$. When $q=2$, $f(\alpha(2))=2\alpha(2)-D_2$.

From equation (14) $\alpha_i \simeq \ln p_i / \ln(l_0 / L)$. The $\alpha$ value distribution for a multifractal is characterized by the relationship $n(\alpha) \simeq (l_0 / L)^{-f(\alpha)} = \exp[-f(\alpha)\ln(l_0 / L)]$.

When estimating the function $f(\alpha)$ in the vicinity of its maximum at $\alpha_0$ using a parabolic approximation:

$$f(\alpha) \simeq D_0 - \eta(\alpha - \alpha_0)^2; \quad \eta = f''(\alpha_0)/2; \quad n(\alpha) \sim \exp\{-\ln(L / l_0)(\alpha - \alpha_0)^2\}.$$

Since $\alpha_i \simeq \ln p_i / \ln(l_0 / L)$ then, the probability distribution function $p_i$ corresponds to

$$P(p) \sim \exp\{-\eta \ln(L / l_0)(\frac{\ln p}{\ln(L / l_0)} + \alpha_0)^2\}. \tag{18}$$

This describes a log-normal distribution where probabilities, denoted as $p_i$, represent the spatial distribution of neutrons, their density, and reactor power (with the index $i$ serving as the coordinate). Equation (18) demonstrates that neutron density within the critical region follows a log-normal distribution across the reactor's volume. The thermal neutron flux density along the radius of a cylindrical reactor core is modeled using a Bessel function of the first kind and zero order. Near the relatively thin boundary layer, the distribution curve shows a slight rise, which, under certain approximations, can be represented by the right tail of the log-normal distribution. Approximation (18) is quite basic and does not account for the reactor's geometry. Thus, it is more accurate to view expression (18) as a parabolic approximation of the spectral function $f(\alpha)$. This relation can also be interpreted differently. The quantity $\alpha$ (16) is proportional to the value $\sum_{i=1}^{N} p_i^q \ln p_i$ that, when $q = 1$, aligns with the entropy described in equation (10

From the model function $\Phi_q$ in expression (2), we transition to a more realistic scenario where the relationship $k_{ef}=p\bar{v}$ holds true. In this context, the probabilities of generating $i = 0, 1, ..., 7$ secondary neutrons during the fission of $^{235}U$ by thermal neutrons are as follows: $\pi_0 = 0.0333$, $\pi_1 = 0.1745$, $\pi_2 = 0,3349$, $\pi_3 = 0,3028$, $\pi_4 = 0,1231$, $\pi_5 = 0,0281$, $\pi_6 = 0,0032$, $\pi_7 = 0,0001$ [15]. Since the fission process produces two fragments (precursor nuclei), each of these fragments, after a certain time $T$ (approximately $10^3$ times longer than a neutron's lifetime), can emit a delayed neutron. To reflect this, we introduce the probabilities $r_0(T)$, $r_1(T)$, and $r_2(T)$, which correspond to the likelihood that one fission results in $i = 0, 1,$ or $2$ delayed neutrons being produced, $\sum_{i=0}^{2} r_i = 1$.

The $T$ value represents the half-life of the precursor nuclei. There are six groups of delayed neutrons categorized based on their delay times. In this study, kinetics is disregarded, and the relationship between $r_i(T)$ and the period $T$ is not considered. The $\Phi_q$ value from expression (2) is substituted by a more intricate function, expressed in the form:

$$\Phi_q = (1-\beta)^q [p^q \sum_{i=0}^{7} \pi_i^q + (1-p)^q] + \beta^q [p^q \sum_{i=0}^{2} r_i^q(T) + (1-p)^q], \quad (19)$$

where for $^{235}U$, $\beta = 0.0064$. Instead of equations (8) - (10) we get:

$$D_0 = \ln \Phi_{q=0} / \ln(l/L) \simeq 0557, \quad \Phi_{q=0} = 13, \quad l_0/L \approx 10^{-2}.$$

$$-S = (1-\beta)[p\sum_{i=0}^{7} \pi_i \ln(p\pi_i) + \ln(1-\beta)] + \beta[p\sum_{i=0}^{2} r_i(T)\ln(pr_i) + \ln(\beta)] + (1-p)\ln(1-p), \quad D_1 = 0.294$$

,

$$D_2 = \frac{1}{\ln(l_0/L)} \ln\{(1-\beta)^2 [p^2 \sum_{i=0}^{7} \pi_i^2 + (1-p)^2] + \beta^2 [p^2 \sum_{i=0}^{2} r_i(T) + (1-p)^2]\}, \quad D_2 = 0.211.$$

The magnitudes of $D_{max} = D_{-\infty} = \alpha_{max}$ and $D_{max} = D_{-\infty} = \alpha_{max}$ (15) rely on the minimal and maximal values of the products $(1-\beta)p\pi_i, i=0,...,7, \beta r_i, i=0,1,2, (1-\beta)(1-p), \beta(1-p)$, where values $r_i, i=0, 1, 2$, not defined. If you set $r_2=r_0=0.1$, $r_1=0.8$, then the minimum value is $(1-\beta)p\pi_7$, and $D_{-\infty} = \frac{1}{\ln(l_0/L)} \ln[(1-\beta)p\pi_7] \approx 2.1$. The utmost combination in this instance will be $(1-\beta)(1-p)$, and $D_{\infty} = \frac{1}{\ln(l_0/L)} \ln[(1-\beta)(1-p)] \approx 0.125$.

From equations (3) - (8) and (19) we construct a spectrum of generalized dimensions (Fig. 2).

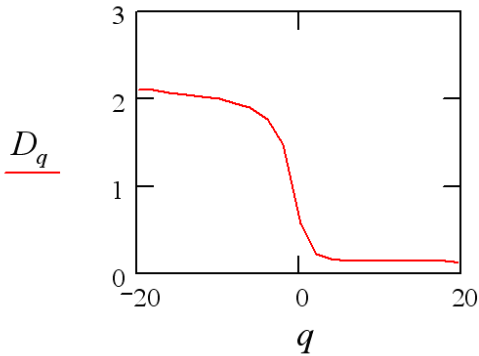

Fig. 2. Range of generalized dimensions for fission chains in nuclear reactors, considering delayed neutrons.

As above, the values are derived from formula (19) $\frac{d\Phi_q}{dq}\big|_{q=0}$, $\frac{d^2\Phi_q}{dq^2}\big|_{q=0}$ and

$$\alpha_0 = \alpha(q=0) = \frac{1}{\ln(l_0/L)}\frac{1}{13}[9\ln(1-\beta) + 8\ln p + \sum_{i=0}^{7}\ln\pi_i + \ln(1-p) + 4\ln\beta + 3\ln p + \sum_{i=0}^{2}\ln r_i + \ln(1-p)].$$

In the expression for $\eta$ from equation (18), the value is worthy

$$D''|_{q=o} = \frac{1}{\ln(l_0/L)}[-2\ln 13 - \frac{2}{13}\frac{d\Phi_q}{dq}\big|_{q=o} + (\frac{1}{13}\frac{d\Phi_q}{dq}\big|_{q=o})^2 - \frac{1}{13}\frac{d^2\Phi_q}{dq^2}\big|_{q=o}],$$

$$\frac{d\Phi_q}{dq}\big|_{q=0} = 9\ln(1-\beta) + 8\ln p + \sum_{i=0}^{7}\ln\pi_i + 2\ln(1-p) + 4\ln\beta + 3\ln p + \sum_{i=0}^{2}\ln r_i,$$

$$\frac{d^2\Phi_q}{dq^2}\big|_{q=0} = 9[\ln(1-\beta)]^2 + 2\ln(1-\beta)[8\ln p + \sum_{i=0}^{7}\ln\pi_i + \ln(1-p)] + 4(\ln\beta)^2 + 11(\ln p)^2 +$$

$$+ 2\ln p(\sum_{i=0}^{7}\ln\pi_i + \sum_{i=0}^{2}\ln r_i) + \sum_{i=0}^{7}(\ln\pi_i)^2 + 2(\ln(1-p))^2 + 2\ln\beta(3\ln p + \sum_{i=0}^{2}\ln r_i + \ln(1-p)) + \sum_{i=0}^{2}(\ln r_i)^2.$$

Within the log-normal distribution of the species (18), assortments of values manifest as likelihood $p$:
$(1-\beta)p\pi_i$, $i=0,...,7$, $\beta pr_i$, $i=0,1,2$, $(1-\beta)(1-p)$, $\beta(1-p)$.

Should we construct the relationship of the multifractal spectrum $f(\alpha)$ from formula (16), employing the method of [1, 25 18], with a measure of the multiplicative ensemble, the resulting dependence is shown in Fig. 3.

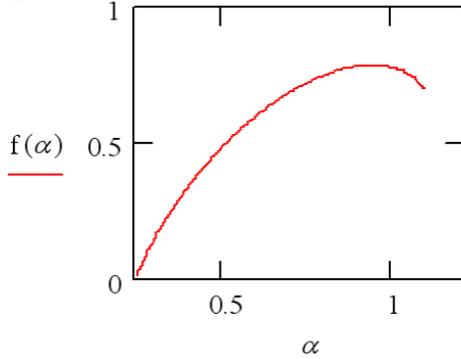

Fig. 3. Multifractal spectrum function for fission chains in nuclear reactors considering delayed neutrons. The function $f(\alpha)$ for the inhomogeneous Sierpinski triangle possesses a similar form [2].

### 3. Implementation of the concept of hierarchical subordination to chain reactions in a nuclear reactor

According to the theory of hierarchical arrangements, a parallel has been drawn between the kinetics for the count of neutrons derived from the theory of proliferation, the neutron quantity of the *n*-th cycle, the tally of elements at the *n*-th tier of the structure, the pace of shift in the likelihood of a chain response, the nature of intensity and vigor of the structural linkage, the measure of device subcriticality, and neutron paths with in the device. Alink has been discovered between the likelihoods of creating as pecific cycle of the neutron's um and the probability of a self-supporting nuclear fission chain reaction happening. It is demonstrated that the Tsallis and Rényi distributions characterizing these events are linked by equations of deformed mathematics, and given certain circumstances can correspond with respect to one another.

#### 3.1. Connection between neutron reactor states and neutron paths.
For neutron processes in a nuclear reactor, the principal attributes are the percolation likelihood, interpreted as the chance of a self-sustaining chain reaction, and the percolation threshold magnitude, which is proportional to the neutron multiplication factor. In [30], are cursive formula was derived for the likelihood of percolation from a source vertex, the chance that a connected segment of the arrangement containing the source vertex (some initial point of the first neutron's appearance in the setup, which initiated a chain reaction), attains the opposing boundaries of the system. Conventionally, mathematically, the scale of the system and the connected segment end to ward infinity, although actual systems are limited. In [30], the value *P(n,c)* signifies the likelihood of percolation from the source vertex to a span *n*. The value *n* in our case is understood as the count of neutron generations in a chain reaction. The figure $c_{c\infty}=inf\{c: P(c)>0\}$ is termed the percolation threshold in [30]. In [1], this value is called the crucial probability where a cluster first emerges, spanning the whole grid. Here $P(c)=lim_{n\to\infty}P(n,c)$. In [7, 18], for the percolation likelihood *P(n,c)* are cursive formula of the form was utilized:

$$P(n+1,c) = c[1-(1-P(n,c))^s], \quad P(0,c) = c. \tag{20}$$

The probability of nuclear fission by a neutron is given by the equation $c=p=\lambda_f(\lambda_f+\lambda_c)^{-1}$ where *c* equals *p* equals $\lambda_f$ divided by ($\lambda_f$ plus $\lambda_c$). The intensity of neutron death, which means the neutrons are either absorbed by the environment or leave the system, during a very small time *Δt* is represented as $\lambda_c$ times *Δt* plus a term that becomes negligible as *Δt* approaches zero ($\lambda_c \Delta t+0(\Delta t)$). The intensity of nuclear fission by a neutron is $\lambda_f$ times *Δt* plus a negligible term ($\lambda_f \Delta t+0(\Delta t)$). Here, $\lambda_f$ is equal to *v* multiplied by $\Sigma_f$, where *v* is the neutron speed and $\Sigma_f$ is the macroscopic fission cross section ($\lambda_f=v\Sigma_f$). The value $s=\bar{v}$, $\bar{v}$ represents the average number of secondary neutrons produced in one fission event. The effective neutron multiplication factor $k_{ef}$ is equal to $k_{ef}=p\bar{v}$. The probability *c=p*, as given by equation (20), is linked to an important value called the percolation threshold, which is connected to the critical point of the reactor. The equations (20), as mentioned in [18], help in determining the critical point of the reactor,

$$P_{n-1} = P_n + N_n^{-1}w(P_n), \tag{21}$$

where $P_n$ is the strength of a structured group at level *n*, and in a random system, this becomes a probability density, it shows the chance of forming a group of different levels, an *n*-level structured group, *w* is how connected the objects are at a certain level, which are like tree nodes, and $N_n$ is the number of nodes at that level. The connection level *w* of objects at a certain level depends on the number of steps *n* to the common ancestor, which shows the distance in an ultrametric space. In our case, the number *n* stands for the number of generations of neutrons in a fission chain reaction. The value *w* represents family ties in a genealogy. By comparing expressions (20) and (21), we find that

$$w(P_n) = N_n[1-P_n-(1-P_n/c)^{1/s}], \quad s = \bar{v}. \tag{22}$$

In [30] and [18], the value *P(n,c)* stands for the chance that a path exists from the starting point to a distance of *n*. In [18], this chance is compared with the likelihood of a continuous chain reaction. The number *n*, which represents the number of generations of neutrons in a chain reaction, is related to time by a factor that varies depending on the reactor type. For thermal neutron reactors, each generation of neutrons lasts about 0.1 seconds. For fast neutron reactors, each generation lasts 3 to 7 times less. The value $N_n$, which is the number of nodes at level *n*, matching the number of neutrons in the *n*-th generation, for a regular tree (Fig. 1a) is equal to

$$N_n = \bar{v}^n. \tag{23}$$

The main feature of hierarchical systems is that they have a self-similar structure [7]. Let's look at the degree of hierarchical connection, which is given by $w(P_n)$ (22), for small values of the argument. When expanding this quantity as a series in the region where $P_n\to 0$ is close to zero, we consider the value $P_n\to 0$, $P_{n0}<P_n$. We find the maximum term of the expansion, which is equal to

$$w(P_n)=N_n A P_n^{1/s}, \tag{24}$$

where $A=P_{n0}^{(s-1)/s}(1-P_{n0}/c)^{(1/s)-1}/c$, $P_{n0}$ is some fixed value of $P_n$, close to 0.

If we compare (24) with the expression from [7] for the case *n>>1* when $P_{n-1}\sim P_n$,

$$w(P)=WP^\beta, \quad P\to 0, \tag{25}$$

where *W=w(1)* is a positive constant, *β=1-D, D≤1* is the fractal dimension of a self-similar object such as an indented coastline [31, 8], we obtain that *1/s= β*, *D=1-1/s=ln $\bar{v}$ /lnq⁻¹*, *q<1* is the similarity parameter, and $P_n \sim q^n$, the connection function satisfies the homogeneity condition $w(qP)=q^\beta w(P)$. We find that at $\bar{v}=2.4$, $lnq^{-1}=ln\bar{v}/(1-1/\bar{v})\approx 1.5$, $q=(\bar{v})^{-\frac{1}{1-1/\bar{v}}}$. From a comparison of (24) and (25), since $W=w(1)=1-c$, we also obtain that $P_{n_0} = [(1-c)c]^{\frac{\bar{v}}{\bar{v}-1}}[(N_{n_0})^{\frac{\bar{v}}{\bar{v}-1}} + \frac{1}{c}[(1-c)c]^{\frac{\bar{v}}{\bar{v}-1}}]^{-1}$. Assuming in equalities (21), (23) that for arbitrary values of $P_n$ the scaling relation $P_n=x_n q^n=x_n s^{-n/D}$ is satisfied, we arrive at the recurrent equality for the function $x_n$:

$$x_{n-1}=\Phi(x_n), \qquad \Phi(x)=q(x+Wx^{1-D}). \qquad (26)$$

The mapping $\Phi(x)$ has two stationary points corresponding to the condition $x=\Phi(x)$: stable $x_s=0$ and critical

$$x_c=(W/(q^{-1}-1))^{1/D}, \qquad q=s^{-1/D}. \qquad (27)$$

The system's behavior is shown through homogeneous functions

$$P_n=x_c s^{-n/D}; \qquad w_n=W^{1/D}(q^{-1}-1)^{-\Delta s - \Delta n}, \qquad (28)$$

where $\Delta=(1-D)/D$, which is the decrement that determines the scale of the hierarchical connection in ultrametric space [7, 8], and this takes into account the vertices of hierarchical trees.

In [7], the continuous limit $n\to\infty$ is employed, the finite difference $P_n - P_{n-1}$ is substituted with the derivative $dP_n/dn$, and an equation of the form (21) is expressed in a continuous format. A comparison between the precise numerical computation and resolutions of approximate analytical expressions demonstrates their convergence as $n$ grows, with agreement already occurring at $n$ values around 15-20. Let us examine solutions distinctly for various kinds of hierarchical trees exhibiting different behaviors of the function $N_n$, the count of nodes at level $n$, which corresponds to the number of neutrons in the $n$-th generation. For modest values of $P$, in the asymptotic form (25) for a regular tree with $N_n$ in the manner of (23), an explicit resolution of this equation of the form:

$$P=W^{1/(1-D)}[(1-u)+ue^{\zeta-\zeta_0}]^{1/D},$$
$$u=DW^{1/(1-D)}/lns,$$
$$\zeta=(n_0-n)lns,$$
$$\zeta_0=n_0 lns, \quad n\leq n_0,$$
$$w=[(1-u)+ue^{\zeta-\zeta_0}]^\Delta,$$
$$\zeta\leq\zeta_0, \qquad w(\zeta_0)=1, \qquad (29)$$

where $\zeta$ is the separation in ultrametric space, $n_0>>1$ is the overall count of hierarchical strata. The rationale $c$ from (20) enters into (29), (32), (33) via $w$ from (22) and (24) and $W=w(1)$. For a fixed arrangement of a hierarchical structure, a vital function is held by the fractal magnitude $D$, the measure of which dictates the robustness of the hierarchical link $w(\zeta)$. In non-steady systems, the likeness factor $q$ varies with duration, and $D(q)$ also shifts. For such intricate setups as the distribution of multiplying neutrons in a core, the hierarchical bonding is multifractal in character [33].

An important part is the range of values that $q$ takes, where the coupling strength $w_q(\zeta)$ is spread out according to a density $\rho(q)$. The overall coupling force is found by the equation $w(\zeta)=\int_{-\infty}^{\infty}w_q(\zeta)\rho(q)dq$. A formula like equation (29), where the fractal dimension $D(q)$ changes, is used as the kernel $w_q(\zeta)$. How this function behaves for the reactor, as calculated, is shown in Figure 1 from [32] and Figure 2b from [18]. These relations only describe the behavior of the hierarchical system as it approaches the limit $1<<\zeta\leq\zeta_0$.

The resulting pattern shows what kind of behavior the system has overall. To get exact answers, you need to use equations like (20) and (21) with numerical methods, as done in [18]. The way probabilities spread across different levels was studied in [7], and it's shown here that the steady state follows a Tsallis distribution. Also, when you use a distribution that includes the lifetime [33], you can get more varied distributions, such as superstatistics and their extensions [11].

Tsallis distributions are just a special type of superstatistics, and they can be expanded further. The chance of forming a self-similar network, like a self-sustaining chain reaction in nuclear fission, goes up as the value of $n$ decreases, reaching its highest point when $n$ is zero, which is the starting point with just one neutron. Even though the single neutron at the beginning has the highest chance of starting a chain reaction, its actual ability to do so isn't very strong yet. The development of hierarchical structures, as described in [7], is seen as a process similar to diffusion on trees that branch out randomly. The structure of these trees depends on a parameter called the heterogeneity parameter, which shows how complex the system is. Similar to how entropy measures disorder in the arrangement of atoms, complexity measures the disorder in how information or communication flows within a system's hierarchy. However, instead of looking at individual atoms, complexity looks at smaller groups of parts, called sub-ensembles, within the entire system.

Relation (20) is written for the number of nodes $N_n$ at level $n$, which corresponds to the number of neutrons in the $n$-th generation of type (23), $N_n=s^n$, where $s=\bar{V}$ is the branching index of the tree. We now use the above-mentioned proportionality between the number of neutron generations and time. Let us compare the expressions for the number of nodes $N_n$ at level $n$ with the time behavior of the number of neutrons, as determined, for example, from the theory of branching processes [15, 34]. Expression (23) was written in [7] for the case of a regular tree shown in Fig. 1a, and since $n$ is proportional to $t$, it corresponds to the time behavior of the number of neutrons, which has an exponential form $e^{-\alpha t}$, valid outside the critical region [15]. It was shown in [34] that in the critical region the dependence is power-law, $t^a$, which coincides with the power-law approximation of the form

$$N_n=(1+n)^a. \quad (30)$$

In [7], the case of a self-similar irregular tree is shown in Fig. 1d. This matches the power-law relationship found in [18] through numerical studies of the boundaries of the critical region. When the branching of the hierarchical tree becomes greater than the golden ratio, which is approximately $a_+=(5^{1/2}+1)/2\approx1.61803$, the behavior seen in simple statistical systems appears. However, the reduction in complexity as the hierarchical connections become more spread out, which is a feature of complex systems, only happens when the branching is weak, and stays within the range of 1 to $a_+$, $1<a<1.618$. For a degenerate tree, Fig. 1c,

$$N_n=1+(s-1)n\approx sn. \quad (31)$$

This value is close to (30) when $a=1$, and it shows a straight-line relationship with time, reflecting the time behavior at the critical point [15, 34]. It's still unclear if movement along the Fibonacci tree relates to any real physical situation in reactors, as shown in Fig. 1b [7]. In such a case, the fissile nuclei must be of a type where the average number of secondary neutrons produced during fission equals the golden ratio $\tau=(5^{1/2}+1)/2\approx1,61803$.

For a degenerate tree with 12 nodes, when looking at the behavior at the critical point, instead of having an exponential dependence like in equation (29), we get a logarithmic dependence of that form:

$$P=W^{-1/(1-D)}[1-u\ln(1+(s-1)(\zeta_0-\zeta)/\ln s)]^{1/D},$$
$$u=DW^{1/(1-D)}/(s-1),$$
$$\zeta=(n_0-n)\ln s, \quad \zeta_0=n_0\ln s,$$
$$w=[1-u\ln(1+(s-1)(\zeta_0-\zeta)/\ln s)]^\Delta, \quad \zeta\leq\zeta_0. \quad (32)$$

In the case where the tree is irregular but grows in a power-law way with respect to the number of nodes and neutrons (30), the intensity and strength of the hierarchical links also follow a power-law pattern based on the distance $\zeta$ in ultrametric space, and this is proportional to the number of generations of neutrons:

$$P=W^{-1/(1-D)}[1+u(1-\zeta/\zeta_0)^{-(a-1)}]^{1/D},$$
$$u=DW^{1/(1-D)}n_0^{-(a-1)}/(a-1),$$
$$\zeta=(n_0-n)\ln s, \quad \zeta_0=n_0\ln s, \quad n\leq n_0,$$
$$w=[1+u(1-\zeta/\zeta_0)^{-(a-1)}]^\Delta, \quad \zeta\leq\zeta_0. \quad (33)$$

The way probabilities of a chain reaction behave is influenced by how likely the reaction is to happen, how close the system is to a critical state, and the degree of criticality. Based on how close the system is to this critical state, there are three main types of behavior (more accurately, four): subcritical and supercritical, where the behavior follows similar rules (23) and (29) but with opposite signs, critical (30), (33), and the critical point (31), (32). In traditional nuclear reactor theory, only subcritical, supercritical, and the critical point are considered, but in the broader theory of phase transitions, the critical region is important. This is because neutrons don't interact directly, and the classical critical exponents apply to them, just like in a self-consistent field [32]. In a reactor's steady operating state, there are many neutrons, and their number is considered very large. In this case, the critical region becomes just a single point. It's worth noting that the exact form of the expression for $w(P_n)$ is known, and the equation for $P$ in the continuum limit can be solved exactly. However, the integrals involved are complicated, making it hard to write out the function $P$ explicitly.

The critical region exhibits a complex three-tiered structure. According to [34], three distinct modes of critical behavior in nuclear reactors were identified based on the nature of control actions and feedbacks. The boundaries separating these modes were determined, and it was revealed that near the critical point, the temporal behavior follows a power-law relationship. In this context, time is directly proportional to the number of neutron generations, a behavior characteristic of (30), self-similar irregular trees as described in [7]. At the precise critical point, the total neutron count becomes proportional to time (31), corresponding to a degenerate tree structure. Neutron trajectories are influenced by the probability parameter $c$ and the multiplication factor, varying according to these factors. In subcritical and supercritical regions, neutrons move along regular tree configurations. However, in the critical region, their paths follow self-similar irregular trees, with the most critical point associated with degenerate trees. Above the critical point, yet within the critical region, self-similar irregular trees again emerge. In the supercritical domain, the movement reverts once more to regular tree structures.

### 3.2. Probabilities of formation of hierarchical levels, distribution by hierarchical levels and by neutron generations.

Self-similar distributions are described by a power law of the form:
$$p(k) \propto k^{-\gamma}. \tag{34}$$

In this context, the exponent $\gamma > 0$ is associated with the degree of the tree vertex, denoted as $k$, which serves as the scale factor in complex networks. Such dependencies are prevalent across systems of different types. Notably, the form of this dependence remains invariant even when the scale of the variable $k$ changes. Replacing $k$ with a scaled value $k/a$, where $a$ is a positive constant, does not alter the distribution pattern described in equation (34). This invariance highlights the robustness of the order distribution for the vertices within the hierarchical tree of a given graph. Division chains ultimately give rise to a hierarchical structure, visually represented by the Cayley tree (refer to Fig. 1).

Typically, the behavior of a hierarchical system is influenced by the cluster structure across all levels. However, the self-similarity property simplifies this process, allowing us to focus only on defining the structure of the smallest cluster and identifying the hierarchical level. A hierarchical tree serves as a geometric representation of ultrametric space, as noted in [35]. Furthermore, [20] demonstrates that describing hierarchical structures can be reduced to analyzing the diffusion process within this space.

The development of intricate hierarchical systems exemplifies anomalous diffusion across various hierarchical levels, ultimately resulting in a stationary distribution that takes the form of a Tsallis or Rényi distribution. Referring to [6], we analyze the probability density $p_\mathrm{u} = p_\mathrm{u}(t)$ of the system's distribution along the ultrametric space coordinates at time $t$. This distribution adheres to the kinetic equation as described in [36, 37]:
$$\tau \dot{p}_\mathrm{u} = \sum_{\mathrm{u}'}(f_{\mathrm{uu}'}p_{\mathrm{u}'} - f_{\mathrm{u}'\mathrm{u}}p_\mathrm{u}). \tag{35}$$

In this context, the dot indicates differentiation with respect to time, $\tau$ denotes the relaxation time, and $f_{\mathrm{uu}'}$ represents the frequency of transitions between hierarchical levels from $\mathrm{u}'$ to $\mathrm{u}$. To analyze the dependence on ultrametric coordinates, imagine a regular hierarchical tree, defined by a constant branching index $s>1$ and a significantly large number of hierarchical levels, $n \gg 1$. Here, the ultrametric coordinate is expressed as an $n$-digit number in a bases-numerical system: $u=u_0 u_1 \ldots u_m \ldots u_{n-1} u_n$, where $u_m = 0, 1 \ldots, s-1$. The transition intensity can be described using a power series expansion $f_{\mathrm{uu}'} = \sum_{m=0}^{n} f(u_m - u_{m'}) s^{n-m}$. The first term of this series $m=0$ corresponds to the behavior associated with the topmost level of the hierarchy, governing the dynamics of the entire system, such as the structure of a branching chain (e.g., a fission process). Conversely, the final term in the series (for *m=n*) represents the dynamics at the lowest or most localized level, associated with the smallest clusters or individual end branches of the chain.

The distance between points $u$ and $u'$ is defined as $0 \leq l \leq n$ if the conditions $u_m = u_{m'}$ for $m = 0, 1, ..., n - (l + 1)$ and for $m = n-l, n-l+1, ..., n$, $u_m \neq u_{m'}$, are satisfied [8, 11]. Consequently, for a fixed distance $l$, the definition implies that the first $n - l$ terms of the given series are equal to zero, while the last $l$ terms include the factor $s^{n-m}$. For $s > 1$, the value of $s^{n-m}$ is significantly smaller than the factor $s^l$, which is the first term among the remaining ones. Therefore, within this series, only the term with $m=n-l$ and the corresponding factor $f_{uu'} \sim s^l = s^{n-m}$ is of notable importance. Similarly, it can be demonstrated that the probability density can be approximated as $p_u \sim s^{n-l} = s^m$. In the case of a random tree, the branching index $s$ becomes variable, leading to a situation where both the transition frequency $f_{uu'} \to f_{n-m}$ and the probability density $p_u \to p_m$ are expressed in terms of the Mellin transformation [37]:

$$f_{n-m} \equiv \int_0^\infty f(s) s^{n-m} ds, \quad p_m \equiv \int_0^\infty p(s) s^m ds. \tag{36}$$

In this context, $f(s)$ and $p(s)$ are understood as weight functions. Moving from the general coordinates $u = u_0 u_1 ... u_m ... u_{n-1} u_n$ in ultrametric space, where $u_m$ ranges from 0 to $s-1$, we transition to the coordinates defined by the level number, which corresponds to the number of neutron generations discussed in the prior section.

Consequently, the fundamental kinetic equation that describes the probability of forming the *n*-th hierarchical level assumes the following structure:

$$\tau \dot{p}_n = \sum_{m>n} f_{m-n} p_n - \sum_{m<n} f_{n-m} p_m. \tag{37}$$

Unlike expression (35), which describes a continuous ultrametric space, here a discrete representation is utilized, corresponding to hierarchical trees similar to those depicted in Fig. 1. The first term on the right-hand side of equation (37) accounts for the hierarchical connections at a level $n$ with lower levels, where $m>n$, while the second handles connections with upper levels, where $m<n$. Of particular interest is that the right-hand side of equation (37) carries an opposite sign compared to conventional statistical systems [38]. This distinctive feature arises from the fact that autonomous systems inherently establish hierarchical connections spontaneously rather than dismantling them [20].

By expanding the probability $p_m$ from (37) into a series in terms of powers of the difference between $n$ and $m$, and considering the limit as $n \gg 1$, we derive:

$$\tau \dot{p}_n = -D(\partial^2 p_n / \partial n^2) + D_n p_n. \tag{38}$$

The lowest moments $\sum_{m<n}(n-m) f_{n-m} = 0$ and $\sum_{m<n}(n-m)^2 f_{n-m} \equiv 2D$ are considered, and the operator $D_n := \sum_{m>n} f_{m-n} - \sum_{m<n} f_{n-m}$ is responsible for determining the difference in the transition intensities from a specific level to both lower and higher levels. In the absence of hierarchy, no conditions $m>n$, $m<n$ arise from equation (37), and the operator has no effect, $D_n = 0$. However, in hierarchical systems, the intensity of transitions is strongly influenced by whether they progress upward or downward within the hierarchical structure. Additionally, we proceed under the assumption regarding the form of the function $D_n$:

$$D_n := -q d p_n^{q-1} \partial / \partial n, \tag{39}$$

where $q$, $d$, are positive parameters. The underlying assumption is that, up to a factor $-d(q-1)$, the integral $\int_n^{qn} D_n p_n dn$ simplifies to the Jackson derivative:

$$D_n p_n^q := \frac{p_{qn}^q - p_n^q}{q-1}, \tag{40}$$

serving as an example of self-similar hierarchical systems [23]. Consequently, the control equation (38) adopts its final form:

$$\tau \dot{p}_n = -(\partial / \partial n)\left(dp_n^q + D_n(\partial p_n / \partial n)\right). \tag{41}$$

The stationary solution of this equation is expressed in the form of the Tsallis distribution [24]:

$$p_n = \left(p_0^{-(q-1)} + \frac{q-1}{\Delta}n\right)^{-1/(q-1)}; \quad p_0 \equiv \left(\frac{2-q}{\Delta}\right)^{\frac{1}{2-q}}, \quad \Delta \equiv D/d. \tag{42}$$

As stated in (42), as the level number $n$ increases, the probability of its formation $p_n$ declines in a power-law fashion, starting from the maximum value $p_0$ associated with the upper level $n=0$.

By incorporating the deformed exponential $\exp_q(x) = [1+(1-q)x]_+^{1/(1-q)}$, $[y]_+ \equiv max(y,0)$, and the effective energy $\varepsilon_n = \left(\frac{2-q}{\Delta}\right)^{\frac{q-1}{2-q}} n$, the probability expressed in (23) adopts the canonical Tsallis framework:

$$p_n = p_0 \exp_q\left(-\frac{\varepsilon_n}{\Delta}\right). \tag{43}$$

According to [39], the effective temperature $\Delta$ complies with standard thermodynamic relations, provided that the distribution across levels in a hierarchical self-similar set is determined by the escort probability $\mathcal{P}_l := \frac{p_l^q}{\sum_l p_l^q}$ rather than the initial distribution $p_l$. References [40] and [7] highlighted that when $q'$ equals $1/q$, the Tsallis escort distribution aligns with the Rényi distribution derived using the maximum entropy principle applied to Rényi entropy. Additionally, [41] demonstrates that Rényi entropy acts as a negative indicator of the extent of conformal transformation in terms of information discrepancy (divergence). The effective temperature is further connected to the probability of nuclear fission as described in equation (20).

The chance of forming a hierarchy and a self-similar chain, or chain reaction, linked to that hierarchy goes up steadily as the value of $n$ decreases. A calculation [7] shows that when the dispersion $\Delta=D/d$, which is the ratio of the diffusion coefficient $D$ to the energy $d$, increases, it greatly affects how the stationary probability spreads across different levels of the hierarchy. When $\Delta$ is much greater than 1, the distribution (42) becomes almost like an exponential distribution at higher levels of $n$.

When we compare distribution (42) with distribution (33), we notice that one acts like an escort of the other when the value of $a$ is zero. These are the Tsallis and Rényi distributions. As mentioned earlier, these distributions become the same if we replace the physical deformation parameter $Q=2-q$ with the value $2-q'$, where $q'=1/q$. In this situation, the spread or variation $\Delta$ of distribution (42) and the parameter $W$ in expression (33) are connected by the relationship $\Delta=1/W$. The fractal dimension of the ultrametric space in (33) is given by the deformation parameter $Q$:

$$D = q' - 1 = (Q-1)/(2-Q).$$

In the general case $a \neq 0$, the relationships involved are of a specific type. It is shown in [6] that the probabilities $P_n$ of different levels in a hierarchical structure and the probabilities of forming each level $p_n$ are connected through expressions known as deformed algebra, which uses a special exponent called $q$:

$$\ln_q P_n = \sum_{m=0}^n \ln_q p_n,$$

$$\ln_q x = \frac{x^{1-q}-1}{1-q}, \quad P_n = p_0 \otimes_q p_q \otimes_q \ldots \otimes_q p_n,$$

$$x \otimes_q y = [x^{1-q} + y^{1-q} - 1]_+^{\frac{1}{1-q}},$$

$$P_n = \exp_q\left(\frac{\sum_{m=0}^{n} p_m^{1-q} - (n+1)}{1-q}\right) = (\sum_{m=0}^{n} p_m^{1-q} - n)_+^{\frac{1}{1-q}},$$

$$P_{n-1}^{1-q} - P_n^{1-q} = 1 - p_n^{1-q}.$$

The non-stationary case was studied in [7, 8] using a self-similar approach. In this case, the system's behavior is controlled by a power-law relationship $n_c(t)$ that describes the characteristic hierarchy scale, such as the number of generations where a percolation phase transition happens or the critical point of a reactor. The probability distribution is shown as a homogeneous function $p_n(t) = n_c^\alpha(t)\pi(n/n_c)$. Since in our situation $n \sim t$ is proportional to $t$, the importance of the self-similar regime depends on the type of $p_n(t)$.

### 4. Percolation threshold and critical point of a nuclear reactor

Neutrons in a nuclear reactor follow paths resembling Cayley trees, which are linked to branching random processes. The percolation probability—defined as the likelihood of achieving a configuration in the Bethe lattice where a continuous path exists through adjacent conductive nodes across the entire lattice—reflects the probability of sustaining a fission chain reaction. When this probability reaches a critical value, a (conditionally) infinite cluster of neutrons emerges. Percolation probability, influenced by both the reactor's operational duration and its size, is tied directly to the reactor's criticality. The temporal evolution of the neutron multiplication factor is analyzed, with particular focus on the initial stages of forming a self-sustaining nuclear fission chain reaction. The study also explores methods for delineating the boundaries of the critical region.

The theory of percolation, originally stemming from the study of liquid or gas flow through random labyrinths, has evolved into a comprehensive mathematical field with numerous physical applications, such as magnetization, conductivity, and other system properties [1, 17, 31, 45-47]. This theory examines the formation of infinite connected structures, or clusters, composed of individual elements. Percolation is defined as the state when there exists at least one continuous path through adjacent conducting nodes that spans the entire lattice. The collection of elements enabling this connectivity is referred to as the percolation cluster. Fundamentally, percolation theory addresses the emergence of interconnected structures in disordered media. From a mathematical standpoint, it is considered a branch of probability theory dealing with graphs, while from a physics perspective, it represents a type of geometric phase transition.

Percolation phenomena are intrinsically linked to fractality, encompassing the concepts of self-similarity and universality. Fractal models for various systems provide an opportunity to uncover new aspects of phenomena that may seem well understood. Numerous physical systems exhibit fractal and multifractal characteristics. As described in [29], a fractal is a structure composed of parts that, in a certain sense, resemble the entire whole. Fractal properties become particularly evident at the phase transition point, within the critical region. The steady-state functioning of a nuclear reactor (NR) occurs precisely at this critical point, making the application of fractal descriptions highly relevant for understanding and characterizing reactor operations.

In reference [17], the spread of rumors within the percolation model is likened to a chain reaction. The relationships observed in percolation theory [1, 45] are also applicable in the broader context of phase transition theories. Fractal concepts have been used to study diverse phenomena such as highly developed turbulence, inhomogeneous star clusters [27], diffusion-limited aggregation, matter destruction processes, blood structure, and more. The analysis of the physical properties of systems exhibiting fractal structures has prompted advancements in analytical methods within the fractal framework. These advances often involve the use of fractional order equations, as the spatial dimensions take on fractional values. Transitioning to neutron transfer equations with fractional derivatives [28] could hold practical significance for reactor calculations, even though distinguishing sharply between percolation processes and diffusion is not always possible [1]. Reference [48] highlights the need for reevaluating transport processes in percolation clusters, fractal trees, and porous systems to derive accurate transport equations for such systems. In branching fractal structures, "super-slow" transport processes may arise, where a physical property evolves more gradually than its first derivative would suggest. In these cases, the fractional derivative's time index represents the fraction of open channels (or branches) available for percolation. Diffusion dynamics are governed by the random motion of particles, where a diffusing particle can theoretically reach any point within the medium. Conversely, percolation occurs within fractal media, where propagation is confined to a finite region when below the percolation threshold. During diffusion from a source, the resulting diffusion front exhibits a fractal structure. In reference [29], the concept of a "shell" surrounding a percolation cluster is introduced. Based on percolation theory, chain reaction processes within reactors are subsequently examined.

The significance of percolation theory in analyzing neutron processes within a reactor lies in its ability to directly derive key equations, such as the neutron multiplication equation and the equation for the reactor's critical size. These equations effectively interpret the foundational principles of percolation theory, highlighting its utility in understanding neutron behaviors in reactors. The relationship governing the propagation speed of disturbances under local supercriticality is particularly noteworthy and is expected to have practical applications. Additional formulations of percolation theory tailored for reactor systems are also likely to draw considerable interest. This relevance stems from the fact that percolation describes a critical process characterized by the presence of a threshold—a specific critical point. At this threshold, flow occurs along a fractal structure whose geometry is defined by criticality principles. Importantly, the fractal's geometric properties are independent of the medium's microscopic characteristics. Below the critical threshold, kinetic processes such as scattering, absorption, and other neutron interactions are confined to limited regions within phase space. However, at the critical point, the fractal structure dominates, emerging as free energy within the statistical ensemble decreases. Under slow external influences, the system tends toward self-organized criticality [25, 49], giving rise to stationary nonequilibrium states that exhibit chaotic and turbulent behavior. These complex dynamics on fractal structures have been explored using the Lorentz model in prior studies, as detailed in references [6-7].

The kinetics and transfer processes within fractal reactor structures necessitate a focused and comprehensive analysis, as highlighted in prior studies [7-8, 27-28]. Near the critical point, long-range correlation effects emerge, leading to non-Gaussian behavior in kinetic processes, which are influenced by the topological invariants of self-similar fractal sets. The dynamics of transfer processes at the percolation threshold are explored in [27-28]. To model these phenomena, fractional derivative equations are employed, incorporating considerations of memory effects, nonlocality, and intermittency.

Exact self-similarities in nature are relatively uncommon; in other words, fractals with constant dimensions are rare. However, fractals with variable dimensions, known as multifractals, are quite prevalent [1-2].

The article demonstrates that the criticality of a reactor is influenced by more than just the probability of neutron-induced nuclear fission and the number of neutrons released during fission. It also depends on the number of neutron generations, which varies based on both the system's size and time progression. For breeding systems with small dimensions, such as critical assemblies or compact reactors, as well as during the initial start-up phase of a reactor when neutron populations are low, these factors should be carefully considered.

**5. Connection between the theory of percolation on Bethe lattices and the neutron multiplication factor**

Clusters in NR possess the geometric structure of Cayley trees [1,17,31,45]. Mathematical outcomes are attainable for Cayley trees (Fig. 1). The percolation limit on the Cayley tree is below unity. Consequently, states both under and over the percolation threshold may be examined.

A Cayley tree, also known as a Bethe lattice, is a structure without any loops. It starts with a central point, called a node, from which $z$ branches of equal length come out. These form the first layer of the tree. Each end of these branches is also a node. From each of these nodes, $z$-1 new branches grow, which creates

*z(z–1)* nodes in the next layer. This process continues forever, forming an infinite Cayley tree where each node has *z* branches. There is only one path connecting any two nodes. In this case, the randomness in how branches grow needs to be considered. Random graph theory can also be used. Understanding the properties of clusters helps in studying their behavior over time. There is a strong link between fractal patterns and statistical distributions. The processes represented by trees are linked to branching random processes [14], which are used to model neutron behavior in a reactor [15] (Fig. 1). We look at a specific part of the problem that focuses on the size and behavior of clusters - groups of connected nodes. A node refers to a fissile nucleus (or a neutron entering the system, which is the starting point of the tree [50]); a bond represents the path of a neutron. Points where neutrons are absorbed are called hanging ends [50] (nodes with only one connection).

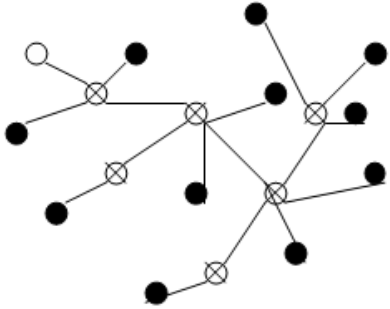

Fig. 4. Paths of neutrons and their offspring in the breeding material: ○ is where the first neutron starts moving; ⊗ shows where nuclei split because of neutrons; ● is where neutrons are absorbed.

If we could track the paths of neutrons inside a nuclear reactor, the observer would notice the special branching pattern of the process that shows how the total number of neutrons changes. Figure 4 shows examples of the paths taken by a single neutron as it moves through a breeding material, including the events like nuclear fission and neutron absorption that cause the number of neutrons to increase or decrease.

In a system characterized by high subcriticality and significantly large negative reactivity $\rho$ values, where $\rho=(k_{eff}-1)/k_{eff}$ and the effective multiplication factor $k_{eff} \ll 1$, the configuration predominantly consists of small clusters with numerous hanging ends. If the rate of neutron loss—either through absorption in the surroundings or escape from the system—within an infinitesimally small time interval $\Delta t$ ($\Delta t \to 0$) is given as $\lambda_c \Delta t + 0(\Delta t)$, and the rate of neutron-induced nuclear fission is represented by $\lambda_f \Delta t + 0(\Delta t)$ (where $\lambda_f = v\Sigma_f$, with $v$ being neutron velocity and $\Sigma_f$ the macroscopic fission cross-section), then the probability of nuclear fission caused by a neutron is described as in Section 3.1:

$$c = p = \lambda_f/(\lambda_f + \lambda_c). \quad (44)$$

The effective neutron multiplication factor, denoted as $k_{eff} = p\bar{v}$, represents the expected number of secondary neutrons $\bar{v}$ generated in a single fission event. As the value of $p$ increases, the sizes of the clusters resulting from chain reactions also grow. When $p$ reaches 1, all the nuclear fuel nuclei in the reactor become interconnected, achieving a maximum $k_{eff\,max} = \bar{v}$. Under such conditions, a critical state is reached that can lead to an explosion. In cases where $1-p$ is much smaller than 1, an infinite cluster is present within the system. There exists a specific critical value, denoted as $p_c$, at which the system transitions from one state to another, marking the initial formation of an infinite cluster. This corresponds to the critical case where $k_{eff}=1$, $c_c=p_c=1/\bar{v}$. In the context of percolation theory, this result has been derived rigorously through mathematical modeling and analysis [1, 17, 31, 29, 45]. The emergence of an infinite cluster signifies a phase transition, as it marks the onset of a self-sustaining chain reaction and represents the system's critical point from the perspective of reactor theory. An essential concept in the study of phase transitions is the order parameter—a physical quantity that plays a central role in describing the transformative processes. Within percolation cluster theory, this order parameter is characterized by $P_\infty$, which represents the probability that a given site on the lattice belongs to an infinite cluster. The behavior of this parameter near the critical threshold, as $p \to p_c$, $p > p_c$, is governed by the relationship or dependence defined at this critical limit:

$$P_\infty = (p - p_c)^\beta. \quad (45)$$

In this context, $\beta$ represents one of the critical indicators, also referred to as scaling indices in percolation theory [1, 45]. The value of $\beta$ dictates the critical behavior of the power of an infinite cluster, denoted as $P\infty$. Within the framework of percolation theory, this probability (45) is often termed the percolation probability, which serves as a fundamental characteristic of a percolation system. The percolation probability is instrumental in describing properties of physical systems reliant on the topology of large clusters, such as spontaneous magnetization or electrical conductivity. Additional parameters, including the average number of nodes in the final cluster, the correlation length $\xi$, and characteristic spatial scales, are also determined. Specifically, at $p < p_c$, these measures relate to the cluster's spatial scale, while for $p > p_c$, they indicate the characteristic size of voids within the system.

The formulas derived from percolation theory for calculating the number of nodes and correlation length bear similarity to equations used in nuclear reactor theory—though obtained differently. Specifically, these equations align with the neutron multiplication formula, $N=(1-k_{eff})^{-1}$, and the critical size equation, $R_{eff}=\pi M(k_{eff}-1)^{-1/2}$, where $R_{eff}$ represents the effective size or geometric parameter, and $M$ denotes the neutron migration length. In this context, the critical threshold is $v = 1/2$. By utilizing concepts from percolation theory and fractal theory, it is possible to establish additional relationships and analyze various properties, such as dynamic critical indices, the cluster skeleton dimension, the spectral (fracton) dimension, and more. Many of these properties have been rigorously derived for Cayley trees as referenced in previous studies [45], [31].

To understand the complex patterns in neutron behavior within reactors, we need to consider how fission happens. By building the multifractal spectrum $f(\alpha)$ [2], using methods from studies [1, 25], and by considering the multiplicative growth of particle numbers, we get the result shown in Figure 3. The shape of this spectrum $f(\alpha)$ for an inhomogeneous Sierpinski triangle is similar [14]. Additionally, the generalized dimension spectrum is also calculated, as shown in Figure 2 [2].

An important value of the percolation threshold is linked to the probability $c=p$ (44). The group of elements that allow flow is known as a percolation cluster. Since it is a connected random graph by nature, its shape can vary depending on how it is set up. Because of this, it is common to describe the size of the entire cluster. The percolation threshold is calculated by dividing the number of elements in the percolation cluster by the total number of elements in the system being studied. Due to the random changes in the state of each element, in a finite system there isn't a clear, exact threshold (the size of the critical cluster), but there is a range of values called the critical range. This range includes the values of the percolation threshold found from different random setups, each with some probability. As the system grows larger, this range becomes smaller and eventually becomes a single point.

Processes on Bethe lattices are usually studied when the lattice is infinite. In this study, we look at a finite lattice, which means there is a limited number of neutrons in the reactor. Considering a limited number of neutrons is important in situations like reactor startup, critical assemblies, and small reactors. The findings from this work can be helpful for fast neutron reactors and for understanding temporary changes in reactor behavior. Besides the probability of percolation and the percolation threshold, there are many other features of the percolation process [45].

We use the notations from formulas (20)-(22). In [30], $P(n,c)$ stands for the chance of percolation from the root vertex to a distance $n$. Here, $n$ refers to the number of generations of neutrons in the chain reaction. The value $c_{c\infty}=\inf\{c: P(c)>0\}$ is called the percolation threshold in [30]. In [1], this same value is referred to as the critical probability where a cluster first appears that covers the whole structure. Here, $P(c)=\lim_{n\to\infty}P(n,c)$, as shown in equation (45). In Figure 5, from [45], the behavior of the function $P(c=p)$ is shown. We will assume that for finite $n$ there is a percolation threshold,

$$c_{cn} = \inf\{c_n : P(n,c) > 0\}. \quad (46)$$

In most situations, the chance of percolation follows the pattern shown in Figure 5 [45]. The recurrence relation for the percolation probability, as found in [30], is given by equation (20):

$$P(n+1,c)=c[1-(1-P(n,c))^s]; \quad P(0,c)=c, \quad (47)$$

where $s=\bar{v}$. In the continuous approximation, there is a derivative of $P(n,c)$ with respect to $c$. To understand how $P(n,c)$ behaves, like through a series expansion or finding the inflection point, you also need to know the second derivative. The formulas for the derivatives, which come from equation (47), are of a certain form:

$$f(n,c)=dP(n,c)/dc\,;\quad f(n+1,c)=1-(1-P(n,c))^{s-1}[1-P(n,c)-csf(n,c)];\quad f(0,c)=1, \qquad (48)$$

$$r(n+1,c)=s(1-P(n,c))^{s-2}[2(1-P(n,c))(dP(n,c)/dc)-c(s-1)(dP(n,c)/dc)^2+c(1-P(n,c))r(n,c)];$$
$$r(n,c)=d^2P(n,c)/dc^2\,;\qquad r(0,c)=0. \qquad (49)$$

From (47), we get an idea of how the percolation probability behaves on the Bethe lattice (Fig. 6), which is different from the behavior shown in Fig. 5, which was for simple lattices. The probability $P(n,c)$ in Fig. 6 is calculated when $n$ equals 750. The vertical line in the figure shows the value of $c_c=\bar{v}^{-1}$ at $n\to\infty$.

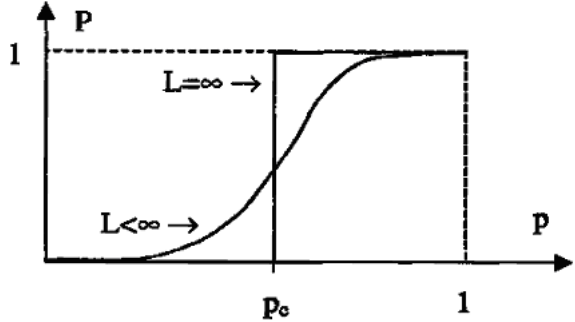

Fig. 5. The probability of percolation occurrence, $P$, depends on the fraction of filled nodes, $p = c$. A smooth curve represents the behavior in a lattice of finite size, while a stepped curve corresponds to an infinitely large lattice.

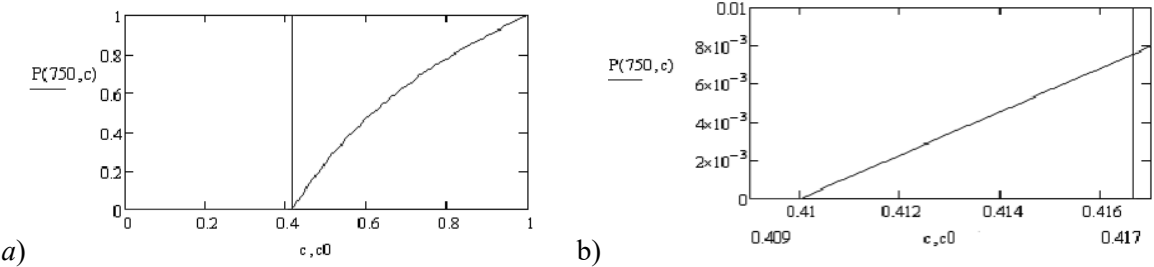

Fig. 6. The probability of percolation for the Bethe lattice is analyzed for $n=750$. It is observed over two distinct ranges of variation: 0 to 1 for case (a), and 0.409 to 0.417 for case (b).

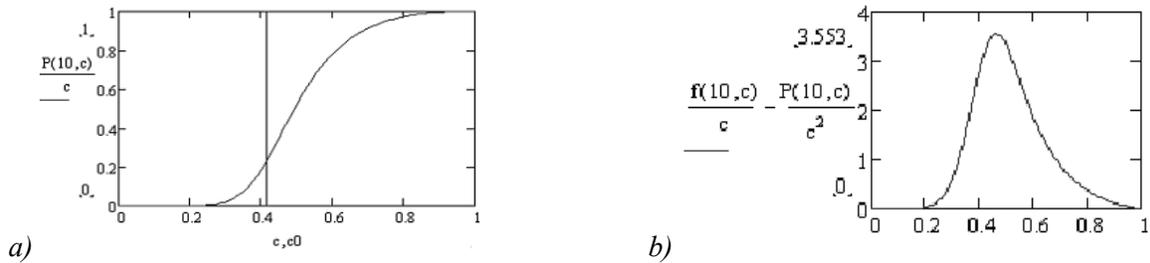

Fig. 7. The behavior of the conditional probability $P(10,c)/c$ (a) and the derivative of the function $P(10,c)/c$ with respect to c (b).

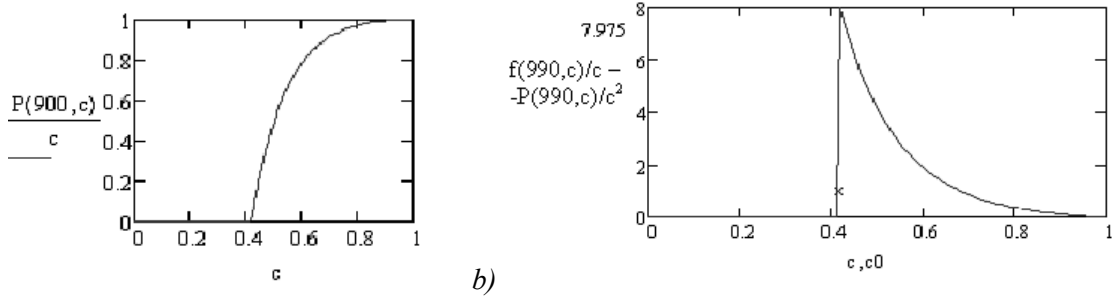

*a)*                          *b)*

Fig. 8. Behavior of the function $P(900,c)/c$ (a) and the derivative of $P(990,c)/c$ with respect to $c$ (b).

In Figures 7 and 8, the behaviors of particular functions and their derivatives are analyzed. Specifically, Figure 7a displays the behavior of the conditional probability function $P(10, c)/c$, while Figure 7b shows the derivative of this function with respect to $c$, denoted as (48). Similarly, Figure 8a illustrates the behavior of the function $P(900, c)/c$, and Figure 8b demonstrates the derivative of $P(990, c)/c$ concerning $c$. Due to the finiteness of the system, where the number of generations is defined as $n = 750$, the critical probability deviates from the theoretical value of $c_c = \bar{\nu}^{-1}$. This deviation is evident in Figure 6b, where the range of $c$ values differs significantly, spanning from 0.409 to 0.417, rather than the full range from 0 to 1 as depicted in Figure 6a. From this observation in Figure 6b, it becomes clear that the critical probability $c_c$ for $n = 750$ is notably lower than that for an infinite lattice where $c_c = \bar{\nu}^{-1}$, with an approximate value of $c_{c750} \approx 0{,}41 < \bar{\nu}^{-1} \approx 0{,}41667$. Therefore, $c_{c750} < \bar{\nu}^{-1}$. In the interval between $c = 0.41$ and $c_{c\infty}$, a non-zero probability exists—ranging between approximately $10^{-3}$ and $8 \times 10^{-3}$—for percolation or, in reactor terms, a self-sustaining chain reaction resulting from uranium nucleus fission. Consequently, it can be concluded that for finite systems, such as this one, both the percolation threshold and the multiplication factor are less than unity.

## 6. Conclusion and discussion

The proposed approach to the theory of nuclear reactors has many aspects and provides many opportunities for a more detailed description. For example, the similarity noted in Fig. 3 between the multifractal spectral function for fission chains in nuclear reactors taking into account delayed neutrons and the same function $f(\alpha)$ for the inhomogeneous Sierpinski triangle. Other aspects of the percolation properties of the neutron population in a nuclear reactor are considered in [51-52]. While this article examines the relationship between the position of the critical point and the number of neutrons during reactor startup or for small neutron systems (for example, neutron assemblies), [51] studies the influence of percolation on feedback in an already operating reactor and Lévy flights, super-diffusion.

The theory of fractal patterns holds significant potential for practical applications in nuclear reactor design and analysis. For instance, consider the findings in [5] regarding the size of the critical region where the percolation threshold values, represented by the critical point $k_{eff}$, can be determined. The width of this region is expressed as $\delta = (l/L)^{1/\nu}$, where $\nu$ represents the correlation radius index corresponding to the reactor's critical size, $R_{cr} = \pi M(k_{eff}-1)^{1/2}$, $\nu = 1/2$, where $M = (L_T + \tau_T)^{1/2}$. Here, $\nu$ is equal to 1/2, and $M$ represents the neutron migration length [26], while $L_T$ is the neutron diffusion length at temperature $T$, $\tau_T$ is the age of thermal neutrons, and $l$ is approximately the size of the lattice, close to the neutron mean free path in the reactor $L \leq R_{cr}$, $L \simeq R_{cr}$. For a nuclear reactor, the value of $\delta$ is reformulated as a function of $\delta = (\pi M / L\sqrt{\bar{\nu}})^2$, where $\bar{\nu}$ represents the expected number of secondary neutrons produced in a single fission event. At the critical point, both $L$ and $R_{cr}$ tend toward infinity, $\delta$ approaches zero, and the critical region effectively collapses into a single point. Furthermore, [17] describes how $\delta$ depends on the total number of lattice nodes $N$ and neutron count. The half-width of the critical point distribution is calculated as $\Delta_N = 2(2\ln 2)\delta(N)$, $\delta(N) = C/N^{1/\nu d}$, $(1/\nu d) = 2/3$, with $C$ being a numerical coefficient. However, in systems with relatively few neutrons—for example, during reactor startup or within a critical assembly—the value of $N$ can remain fairly small. As a result, the critical point can vary within a finite range with a certain probability, leading to fluctuations in the critical point. In [17], researchers quantified the probability that the percolation threshold (corresponding to the reactor's critical point) deviates from the mean value $k_{eff}$=1 by an amount within the interval $\delta$. For large values of $N$, however, this distribution transforms into a sharp peak where all but one percolation threshold value has zero probability. Consequently, the percolation threshold ceases to be random and becomes definitive. The relationships derived in [17] offer valuable insights into reactor behavior as it approaches criticality during startup, which serves as a fundamental factor for ensuring the operational safety of nuclear power plants.

A study presented in [17] examines the fraction of nodes that form part of the skeleton of an infinite cluster—those points that have at least two paths branching out in different directions. For the critical indices relevant to the reactor, the proportion of nodes within the skeleton of an infinite cluster at the critical point is of the same order of magnitude as the proportion of nodes in the entire infinite cluster. This reflects a scenario where the fraction of dead ends within the cluster, representing neutron absorption points, is comparable in magnitude to the total number of nodes in the infinite cluster. This behavior contrasts with many other fractal systems, such as ferromagnets, where the bulk of the infinite cluster's "mass" typically resides in its dead ends.

The kinetics and transfer processes within fractal reactor structures are discussed in [7, 8, 27]. A key issue involves comparing the outcomes derived from equations incorporating fractional derivatives, which are utilized to describe non-stationary processes on fractal structures [27, 28], with those obtained from conventional equations in Euclidean space. As observed in [29], in many comparable scenarios, the differences in results are often negligible.

In some works (for example, in [31]), the value of the critical region is described as $c \approx c_c + B/L + ...$, where $B$ is a constant, $L$ is the size of the system. In [53], it is noted that the size of the Bethe lattice is proportional to $lnN$. Since $N=n$, this corresponds to the expression $1.43/n$. It is possible to estimate the time during which the values of $c_0=c_{c\infty}$ reach a certain given level. For example, the value of $10^{-6}$ is reached in $n=1,43 \cdot 10^6$ neutron generations. For thermal neutron reactors, where the average lifetime of a generation, taking into account delayed neutrons, is $10^{-1}$ s, this time is $1,43 \cdot 10^5$ sec=1.655 days. For fast neutron reactors, where the average lifetime of a neutron generation is $10^{-4}$-$10^{-8}$ sec, this time decreases by 3-7 orders of magnitude. The considered methods for determining the critical point may prove useful during reactor start-up or during its transient processes, when the probability value $c$ changes due to manipulations with the absorbing rods.

This article introduces an innovative approach to analyzing complex processes within a nuclear reactor. It utilizes synergetic techniques combined with fractal and percolation methods for characterizing intricate systems, alongside the theory of hierarchical subordination. These advanced research methods enable a deeper exploration of reactor system behaviors. By comparing them with conventional approaches to studying neutron-nuclear processes in reactors, researchers can identify more nuanced aspects of these phenomena, integrate them into analyses, and enhance reactor safety.

The complexity of hierarchical trees discussed in [42] is defined by the silhouette, represented as $s_l=ln(M_l/M_{l-1})$, where $M_l$ denotes the number of nodes at level $l$. These expressions are outlined in equations (23), (30), and (31). Within a reactor, the ratio $M_l/M_{l-1}$, reflecting the number of neutrons across consecutive generations, serves as an indicator of the neutron multiplication factor. For regular trees, as described in (23), $s_l = ln\bar{\nu}$. At the most critical condition, where $M_l=M_{l-1}$, $s_l$ becomes 0. This situation corresponds to a single neutron being produced per generation, resulting in what is referred to as a degenerate hierarchical tree. For cases involving degenerate trees where $s = \bar{\nu}$:

$$s_l = \ln[1 + \frac{(s-1)}{1+(s-1)(n-1)}] \approx \frac{(s-1)}{1+(s-1)(n-1)}.$$

This value approaches zero either at a specific point $\bar{\nu}=1$ or as a certain parameter increases indefinitely, $n \to \infty$. For self-similar trees, as expressed in equation (30), $s_l=ln(1+1/l)^a \approx a/l$. This value converges to zero as $l$ approaches infinity, as previously noted in [18]. Furthermore, studies in [7, 8] demonstrate that a more accurate representation of the silhouette of a self-similar tree, as well as the neutron multiplication coefficient relevant to this value for a breeding reactor system, is provided by the Jackson derivative described in equation (40).

A powerful approach for analyzing complex systems of this nature is the use of information geometry applied to probability distributions [43, 44]. The Rényi and Tsallis distributions mentioned earlier are derived by employing the maximum entropy principle in conjunction with the respective Rényi $H^{(\alpha)}{}_T(p) = \frac{1}{1-\alpha}(\int p^\alpha(x)d\mu(x)-1)$ and Tsallis $H^{(\alpha)}{}_T(p) = \frac{1}{1-\alpha}(\int p^\alpha(x)d\mu(x)-1)$ entropy formulations. These entropies are associated with Rényi divergence $D^{(\alpha)}{}_R(p|q) = \frac{1}{\alpha-1}\log(\int q(\frac{p}{q})^\alpha d\mu(x))^{1/\alpha}$ and Tsallis-type $D^{(\alpha)}{}_T(p|q) = \frac{1}{1-\alpha}(1-\int p^\alpha q^{1-\alpha}d\mu(x))$ ($p$ and $q$ probability distributions). The generalized Pythagorean theorem [43, 44] is relevant to these measures, providing a framework for expressions related to maximum entropy and other significant physical outcomes. This methodology is critical for conducting a detailed examination of reactor systems.

The way neutrons move and spread in a nuclear reactor, along with how they form complex patterns, shows the complicated processes happening during nuclear fission and neutron movement. Using certain rules from percolation theory that apply to special types of lattices, we can understand how important factors, like the multiplication factor, change over time. These estimates show that getting exactly one multiplication factor can only happen if there are an infinite number of neutron generations, which means an infinite amount of time and an infinitely large reactor system. But for real-world conditions, we can figure out how long it takes to reach very small changes from the one multiplication factor.

These findings are important for reactor safety, especially in situations with fewer neutrons, like when a reactor is starting up or in critical assemblies. They also help in better understanding how reactors behave during sudden changes or temporary situations, transient processes in the reactor.